# Classification of Ion Sources

*R. Scrivens*
CERN, Geneva, Switzerland

**Abstract**
In this chapter, the anatomy of an ion source is briefly described, as well as a few features of particle motion in electric and magnetic fields, and of particle dynamics and plasmas. Using this information, different types of ion sources are described, highlighting their main mode of operation.

## 1 Introduction

The generation of ions for an ion source is based fundamentally on three main processes, which are

i) electron impact ionization,

ii) photo-induced ionization and

iii) surface ionization.

In addition to these, some other atomic and molecular processes can be the process leading to ion extraction, for example, charge exchange between ions and atoms, and molecular dissociation into atoms and ions.

Ion sources have always been soundly rooted in the research arena, which has led to continual optimization of the design for each application type and results in an enormous array of ion source types (just a few of which are shown in Fig. 1). In this chapter we will journey through the main different types of ion source, dealing with the main physical processes required for each source to work. To do this, ion sources will be classified into the following general types:

- electron bombardment,
- d.c./pulsed plasma discharge,
- radio-frequency (RF) discharge,
- microwave and electron cyclotron resonance (ECR),
- laser driven,
- surface and
- charge exchange.

Several excellent books on ion sources exist, but, for the most part, Bernhard Wolf's book *Handbook of Ion Sources* [1] has been used as the main source of information for this chapter.

Overall, this chapter will give a very basic overview of each source type and the processes that make them work. Much more detailed information will be given for most of the source types in the other chapters of these proceedings specializing on each source configuration or process.

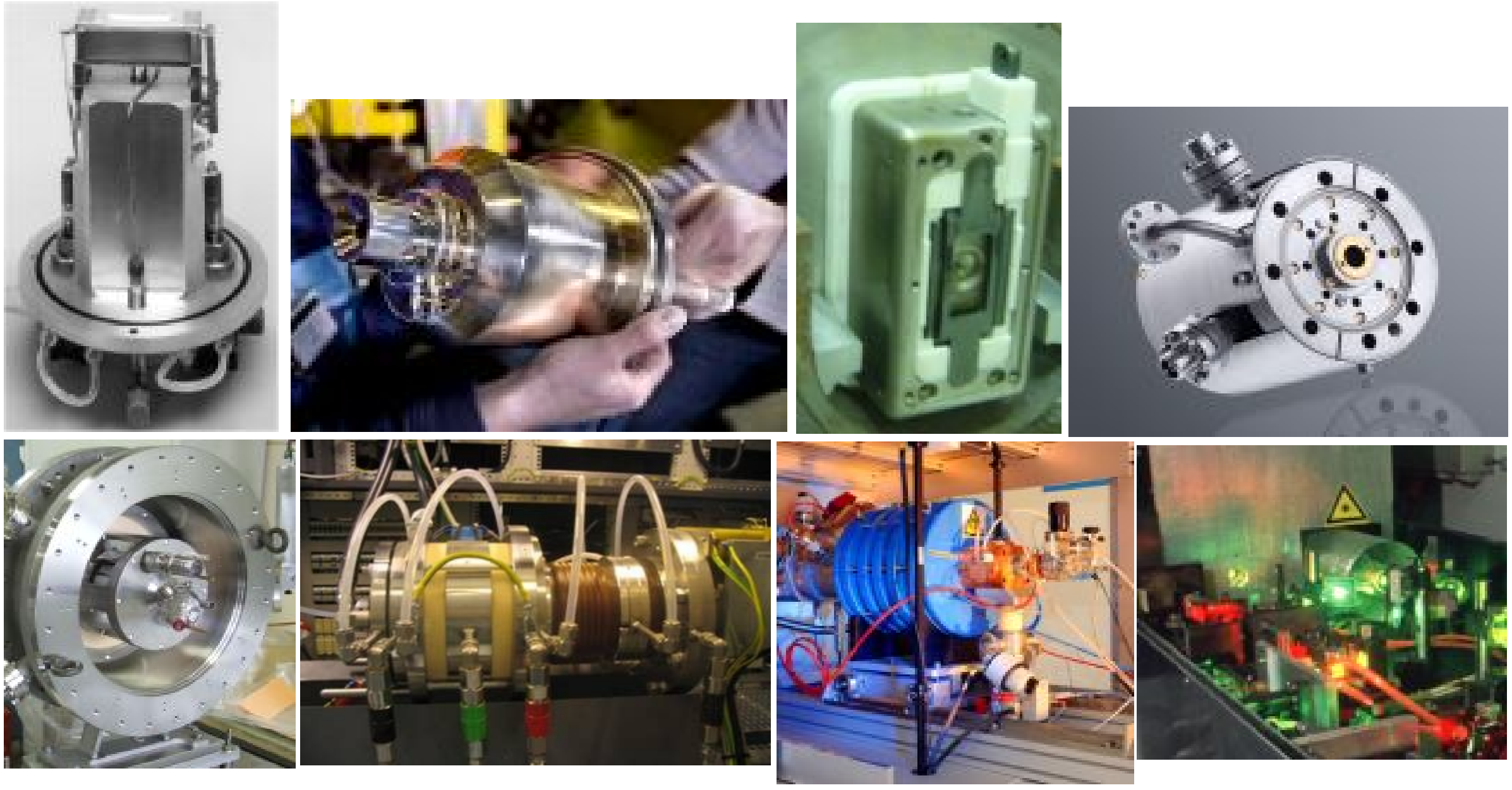

**Fig. 1**: An array of ion sources of various types. Top row: Freeman, Duoplasmatron, Magnetron, Penning. Bottom row: RF ion source, Chordis, ECR, a laser.

## 2 Basic components of an ion source

Before going any further with the description of ion source types, the basic components of an ion source need to be known. Ion sources are so diverse in type that a fully general scheme cannot be given, but for illustration Fig. 2 shows the main components inside a cathode-type ion source, and they can be summarized as follows.

- *Main chamber.* Somewhere in the source there needs to be a chamber in which the ionization processes will take place and through which the created ions will drift in order to arrive at the region for extraction. This chamber will have to be vacuum-tight (or placed in another outer vessel), but depending on the source type it could be metal or ceramic. Somewhere in this chamber there must be a hole through which the ions will exit the source.
- *Material.* For the ion type required, material must be supplied into the source. This can be in the form of gas or a compound gas containing the desired ion type, as well as solid (or even liquid) material that will be heated to produce a gas. Sometimes the material can be introduced into the electrodes, such that it will be sputtered into the main source chamber.
- *Ionization energy source.* The ionization of atoms requires some energy, and therefore power will have to be delivered in some form to the source (for example, via light, or electrical power).
- *Extraction system.* The hole from which the ions will emerge will need to have an electric field applied so that the ions can be extracted from the source, and accelerated to an initial energy.

From these fundamental components onwards, different source types start to diverge dramatically, which will be seen in the next sections.

In addition to the above components, many systems are needed to make the source operational, including vacuum pumping systems, power supplies and controls, to name but a few.

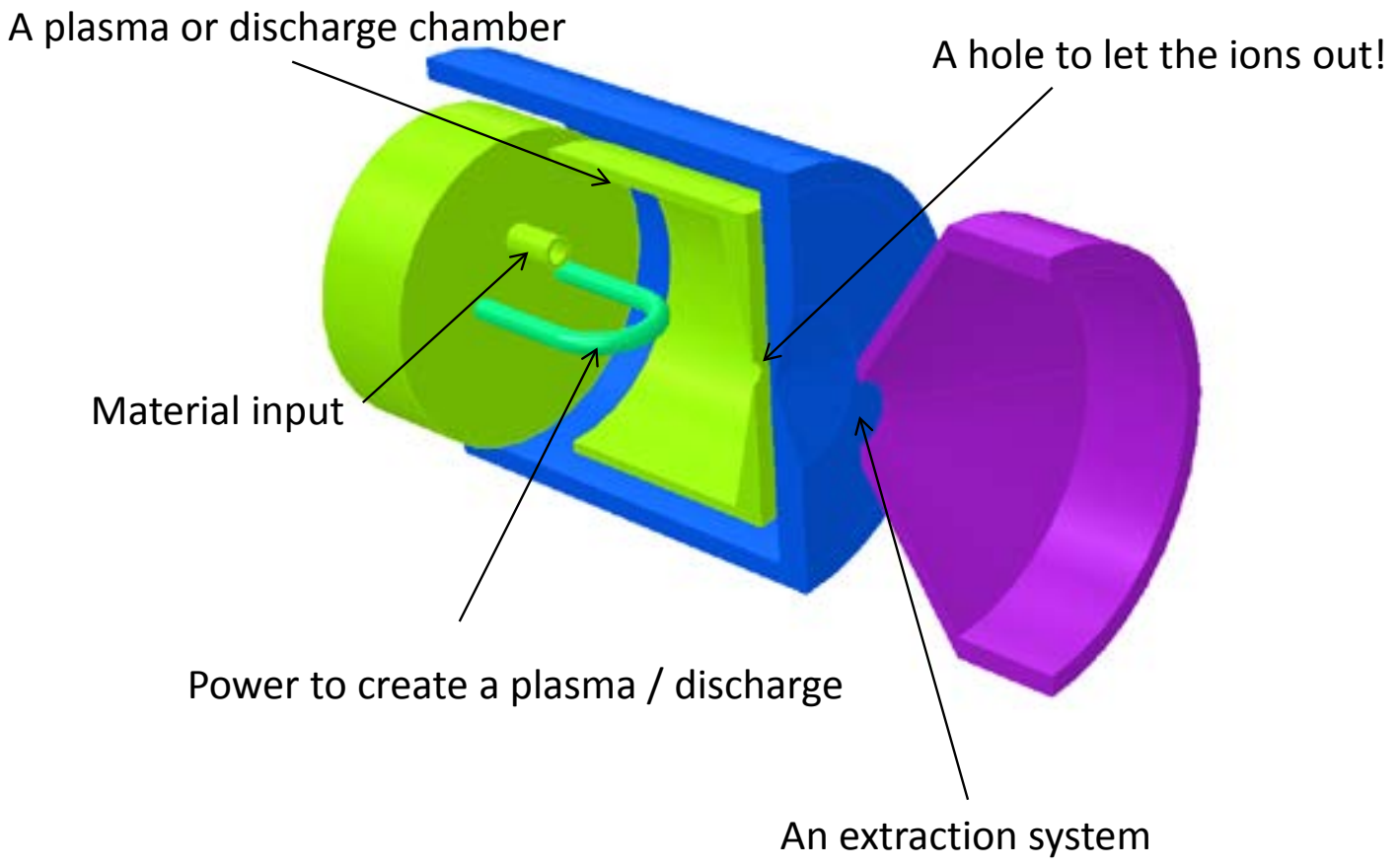

**Fig. 2:** The scheme of a typical ion source (for this example, using electron impact ionization)

## 3 Simple charged-particle dynamics

The charged particles in an ion source, that is, both the positive (and negative) ions and electrons, undergo forces due to the electric and magnetic fields. These fields are both externally applied to the source volume, and generated by the assembly of charged particles.

In order to have a preliminary understanding how the majority of sources are able to control the particles inside them, some simple single-particle dynamics is needed.

### 3.1 Charged particle in a magnetic field

In this simplest situation, a charged particle will make circular orbits in the plane perpendicular to the magnetic field, with a cyclotron radius ($\rho_c$) and a cyclotron frequency ($\omega_c$) given by

$$\rho_c = \frac{\sqrt{2mE_\perp}}{qeB}, \tag{1}$$

$$\omega_c = \frac{qeB}{m}. \tag{2}$$

These equations show that the frequency of oscillation is constant for a given particle, and does not depend on the particle velocity around the magnetic field (as long as the energy is not high enough for relativistic corrections to be needed), but only on the magnetic field strength.

In the direction parallel to the magnetic field, the particle is able to drift without undergoing any force.

### 3.2 *E* × *B* drift

If an electric field is parallel to a magnetic field, a charged particle is accelerated along the direction of the electric field and is not subjected to any force due to the magnetic field.

However, if the electric field and magnetic field are perpendicular, a charged particle is accelerated by the electric field, and as it gains velocity and the magnetic field starts to curve its trajectory. Eventually the trajectory direction starts to oppose the electric field and the particle decelerates until it comes to rest, but displaced in a direction perpendicular to both the electric and magnetic field vectors.

This drift velocity is given vectorally by

$$v_{\mathrm{drift}} = \frac{\boldsymbol{E} \times \boldsymbol{B}}{B^2}, \tag{3}$$

i.e., the drift velocity is independent of the particle charge and mass.

### 3.3 Collisional drift

The charged particles within a plasma undergo long-range elastic collisions that are able to change the particle's trajectory. The rate of these collisions can be characterized by the collisional frequency, given by

$$v_{\mathrm{coll}} = 2 \times 10^{-6} n_{\mathrm{e}} \ln \Lambda \; T_{\mathrm{e}}^{-3/2} \; (m_{\mathrm{e}}/m_{\mathrm{particle}})^{1/2} \; \mathrm{s}^{-1}, \tag{4}$$

where $m_{\mathrm{particle}}$ is the mass of the particle whose collisional frequency is being considered, and $\ln \Lambda$ is the Coulomb logarithm (a factor that describes the difference between long- and short-range collisions, and is typically taken to have a value of 10 for conventional plasmas).

As the charged particles in a plasma make orbits around the magnetic field, with a radius typically of the size of the cyclotron radius given by Eq. (1) (where the particle temperature can be used as an approximation for the energy), the collisions cause the trajectory to be changed. The rate at which the particles drift is a direction perpendicular to the magnetic field ($D$) due to these collisions can be approximated by

$$D \sim \rho_{\mathrm{c}}^2 v_{\mathrm{coll}} \sim \left( \frac{\sqrt{2 m_{\mathrm{particle}} E_{\perp}}}{eB} \right)^2 \frac{1}{T^{3/2}} \left( \frac{m_{\mathrm{e}}}{m_{\mathrm{particle}}} \right)^{1/2} \sim \frac{m_{\mathrm{particle}}^{1/2}}{T^{1/2}}. \tag{5}$$

From this result we see that, when a plasma is confined by a magnetic field, heavier particles drift perpendicularly to the field more quickly, as do particles in cooler plasmas, which is at odds with the conventional kinetic energy equation (i.e., $E = ½mv^2$).

## 4 Ion source types

### 4.1 Electron impact ionization sources

When an electron strikes a target atom (or molecule, or ion) there is a chance that during the impact one or more electrons are removed from the target particle, which turns it into an ion or increases its charge-state. Ion Sources using electron impact ionization work in practice by supplying a source of electrons (typically from a cathode), and a way to accelerate them to an energy sufficient to cause ionization of the material in a chamber. A schematic of an electron impact ionization source is shown in Fig. 3.

Electrons are made available from a cathode in the source, which in most cases will use thermionic emission, i.e., heating the material so that electrons gain some energy and are able to escape the work function of the cathode material (the energy required per electron to leave the material). The rate of electron emission from a surface ($J$, the electron current density) is given by the Richardson equation, Eq. (6), for a temperature $T$ and material surface work function $\phi_{\mathrm{s}}$:

$$J = AT^2 \exp\left( \frac{-e\phi_{\mathrm{s}}}{kT} \right) \tag{6}$$

where $A$ is the Richardson–Dushman constant:

$$A = \frac{4\pi e m_e k^2}{h^3} \approx 1.2 \times 10^6 \ \mathrm{A\,m^{-2}\,K^{-2}}, \tag{7}$$

which, although a constant when the equation is derived from first principles, in practice turns out to be dependent on the material. Values for the work function and the constant $A$ can be found in the literature for many cathode materials, and illustrative calculations of the current density that can be emitted as a function of temperature is shown in Fig. 4.

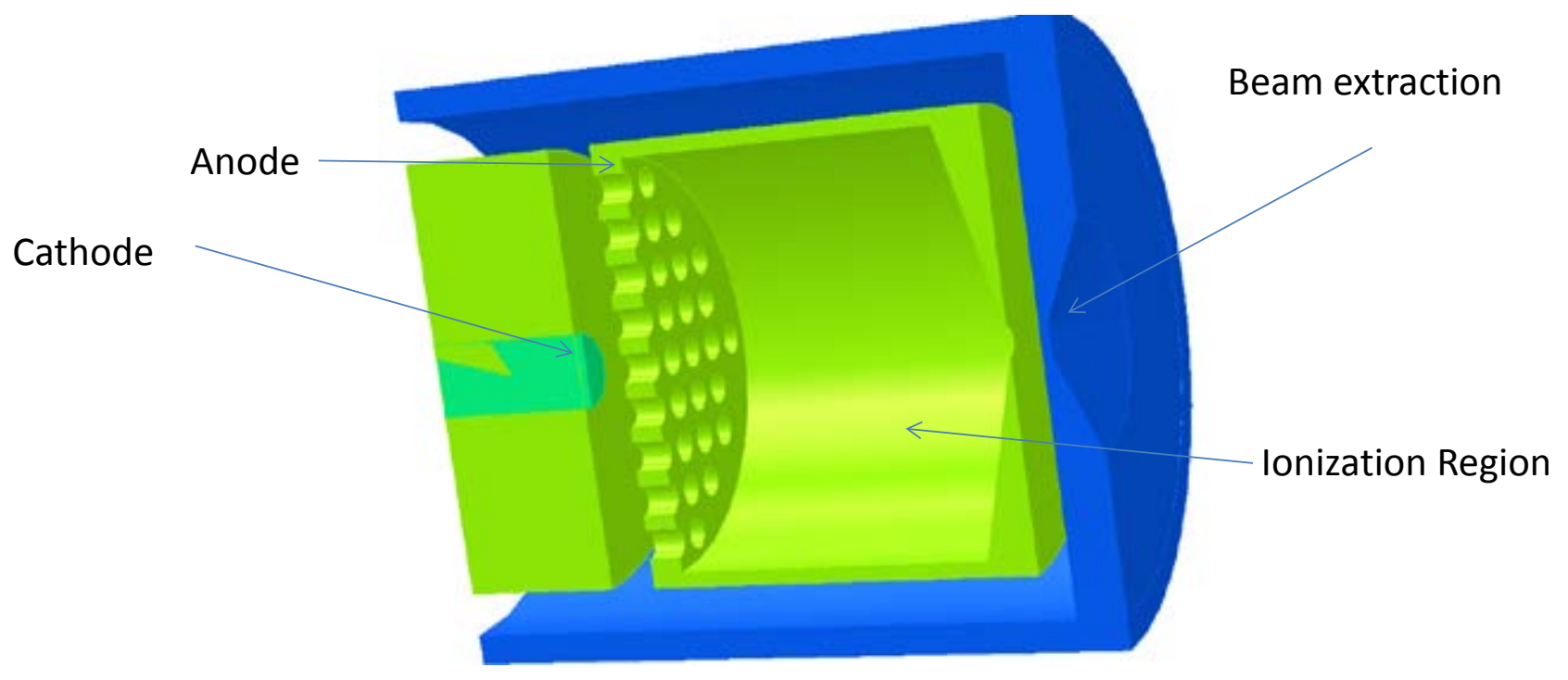

**Fig. 3:** Scheme of an electron impact ion source

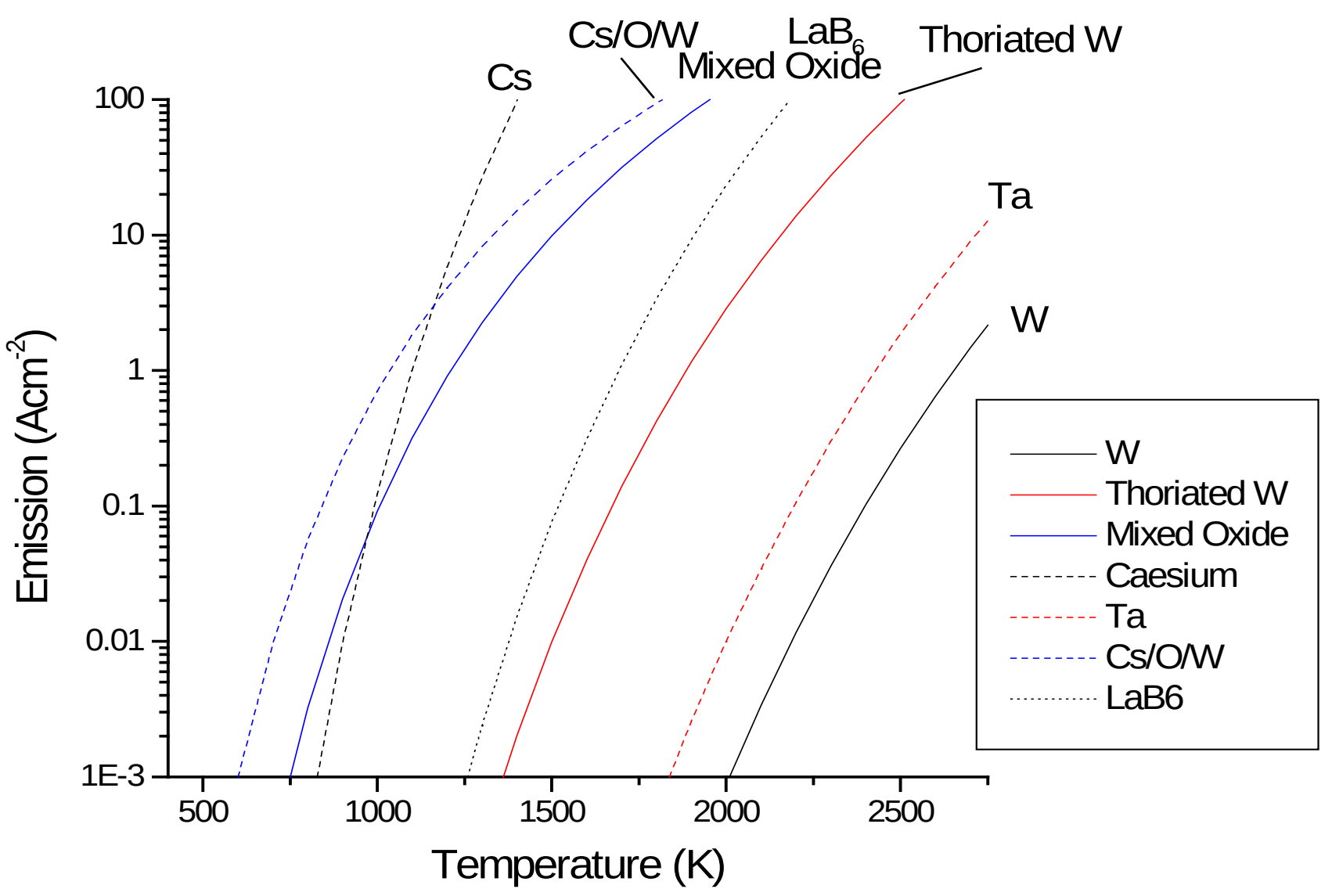

**Fig. 4**: Electron current density emitted from a material as a function of temperature for several cathode materials. Note that the temperature range shown for Cs is not valid, as it is well above the melting point for this element.

The target for the electrons will be neutral atoms or molecules inside the source ionization region. The rate at which ions will be created inside this source depends on the current of electrons available, and the cross-section for ionization of the target particle. We can understand the meaning of the cross-section ($\sigma$) as part of the formula:

$$\sigma n_{\text{atom}} = \frac{1}{L_{\text{collision}}}, \tag{8}$$

where $n_{\text{atom}}$ is the number density of the target particles inside the source volume, and $L_{\text{collision}}$ is therefore the mean path the electron will travel between events causing ionization. If we generate an electron current in a source, and let each electron flow from the cathode to the anode with a distance $L_{\text{path}}$, then by comparing the total distance covered by all electrons to the distance needed to cause an ionization event, the number of ions created per unit time ($\mathrm{d}n_{\text{ion}}/\mathrm{d}t$) can be calculated as

$$\frac{\mathrm{d}n_{\text{ion}}}{\mathrm{d}t} = \frac{I\sigma n_{\text{atom}} L_{\text{path}}}{e}, \tag{9}$$

from which it can be seen that higher electron currents, cross-sections, gas density and path length for electrons to flow are all ways to increase the production of ions. In electron beam impact ionization sources, $L_{\text{collision}} >> L_{\text{path}}$, and therefore the majority of electrons flow from the cathode to the anode without causing an ionization event.

Data for the cross-section have been measured for most ions, and the values for a few of them are shown in Fig. 5. These data clearly show how the ionization cross-section can be very different depending on the ion type, and of course it is necessary that the electron energy is at least higher than the ionization energy.

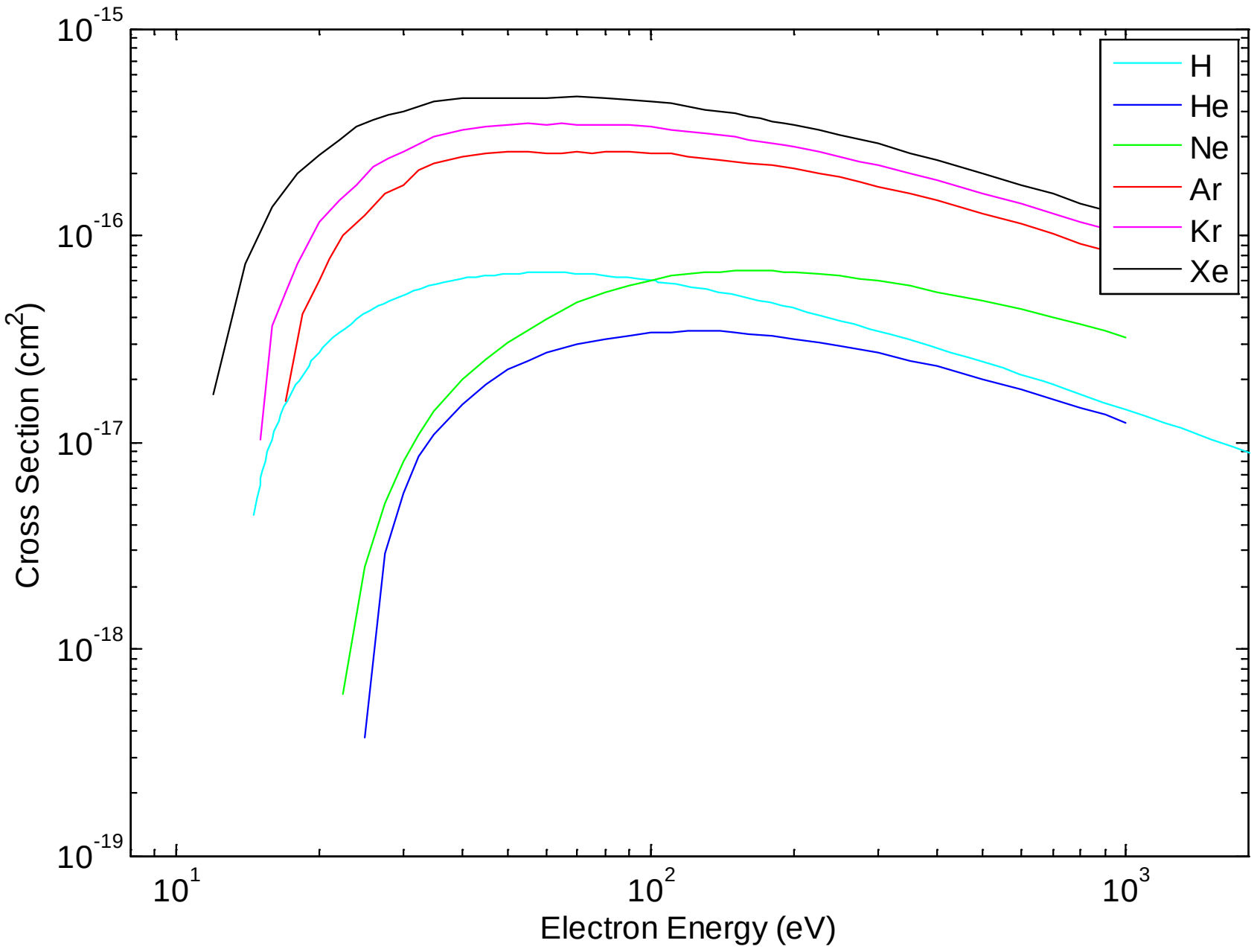

**Fig. 5**: Electron impact ionization cross-section as a function of electron energy, for selected noble gases [2] and hydrogen [3].

The operation mode of the source is therefore to generate electrons from a cathode, and accelerate them to an energy of a few times the ionization potential, often across a short distance to a grid. From there on, the electrons can drift in a field-free region, with the gas. The grid also prevents ions from the ionization region from being accelerated back towards the cathode, but instead have a chance to drift towards the exit hole of the source. A practical example of a FEBIAD (forced electron beam induced arc discharge) used for radioactive isotope ionization can be found in Ref. [4].

### 4.2 Plasma discharge ion sources

For the description of electron impact ionization sources above, the electron density and dynamics were dominated by thermionic emission from the cathode, and by the application of external fields. The charge density in the source was so far not high enough to affect the motion of electrons and ions.

As we move to high densities of neutral particles, electrons and then hopefully ions, several effects take place:

- Higher gas densities mean that $L_{\text{collision}} \sim L_{\text{path}}$ and therefore the majority of electrons will cause an ionization. This means that the electrons lose energy when they cause an ionization, and that new, low-energy electrons are created.
- Large numbers of ions are created, which can be accelerated to the cathode, where they also cause the emission of electrons from the cathode surface.

As the density of ions and electrons in the source increases, a plasma is considered to have been created, which leads to new effects, for example:

- quasi-neutrality (the density of positive and negative charges will generally balance),
- screening of externally applied electric fields (through sheath layers, which are not charge neutral).

The distance over which the screening of fields takes place in a plasma is characterized by the Debye length, which is given by

$$\lambda_{\text{D}} = \sqrt{\frac{\varepsilon_0 k T_{\text{e}}}{n_{\text{e}} e^2}} \, , \tag{10}$$

where $T_{\text{e}}$ is the temperature of the electrons and $n_{\text{e}}$ the electron density. If the Debye length is smaller than the source plasma, then bulk plasma effects will be important, and sheath and screening layers around electrodes will be present. The Debye length as a function of electron density and electron temperature is given in Fig. 6.

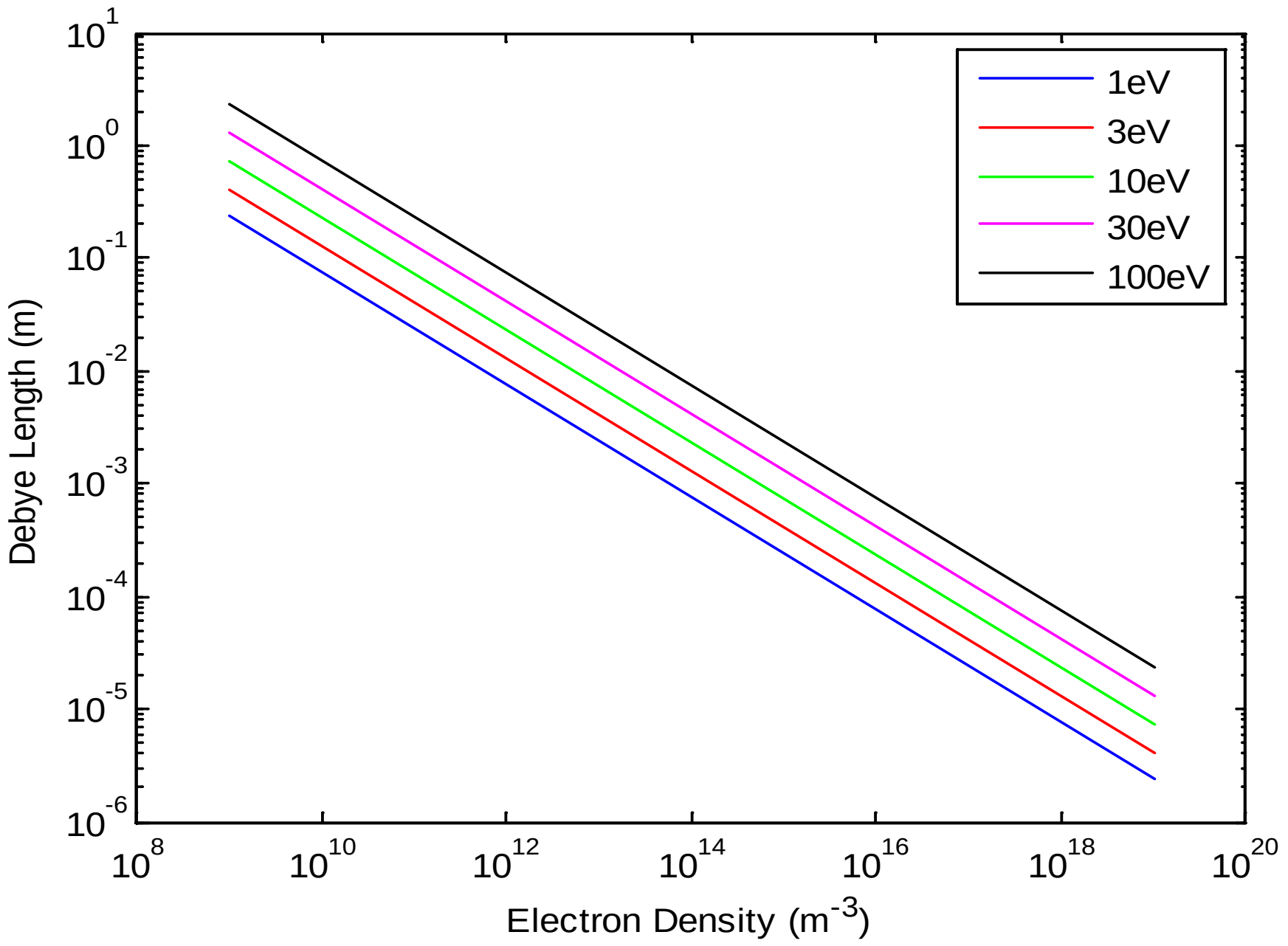

**Fig. 6:** Plasma Debye length as a function of electron density and electron temperature

The creation of significant numbers of ions and electrons from the initial gas will lead to the possibility of sustaining a discharge (i.e., the current flowing between cathode and anode electrodes is no longer dominated by the thermionic electrons from the cathode, but by new ions and electrons created from the gas, as well as secondary electrons from ion impact on the cathode surface).

Paschen investigated the breakdown in gases in the 1880s, and found that the voltage required to produce breakdown in a gas ($V_\mathrm{b}$) was given by

$$V_\mathrm{b} = \frac{aPd}{\ln(Pd) + b}, \tag{11}$$

where $P$ is the pressure, $d$ is the distance between the electrodes, and $a$ and $b$ are gas-dependent constants. For a given gas, the required voltage for a sustainable breakdown is a function of the product of the pressure and distance between the electrodes, and not just the electric field and pressure.

The plot of this function for a gas is called the Paschen curve (Fig. 7), for which we see that there is a minimum voltage needed to break down a gas (for example, in air this is 327 V). The plots also reveal some rather counter-intuitive phenomena. For example, considering a fixed pressure $P$, we can imagine applying a voltage $V$, and changing the distance between our electrodes ($d$). Starting with a very small distance (on the left side of the plot), we are below the Paschen curve, and so breakdown does not occur. As we increase the distance, we move above the curve and a breakdown can be made. Further increase $d$ and we finally reach a distance where the discharge cannot be sustained.

It should be noted that the parameter $b$ is also a function of the number of secondary electrons from the cathode per incident ion, and so Paschen curves are in fact dependent on the cathode characteristics.

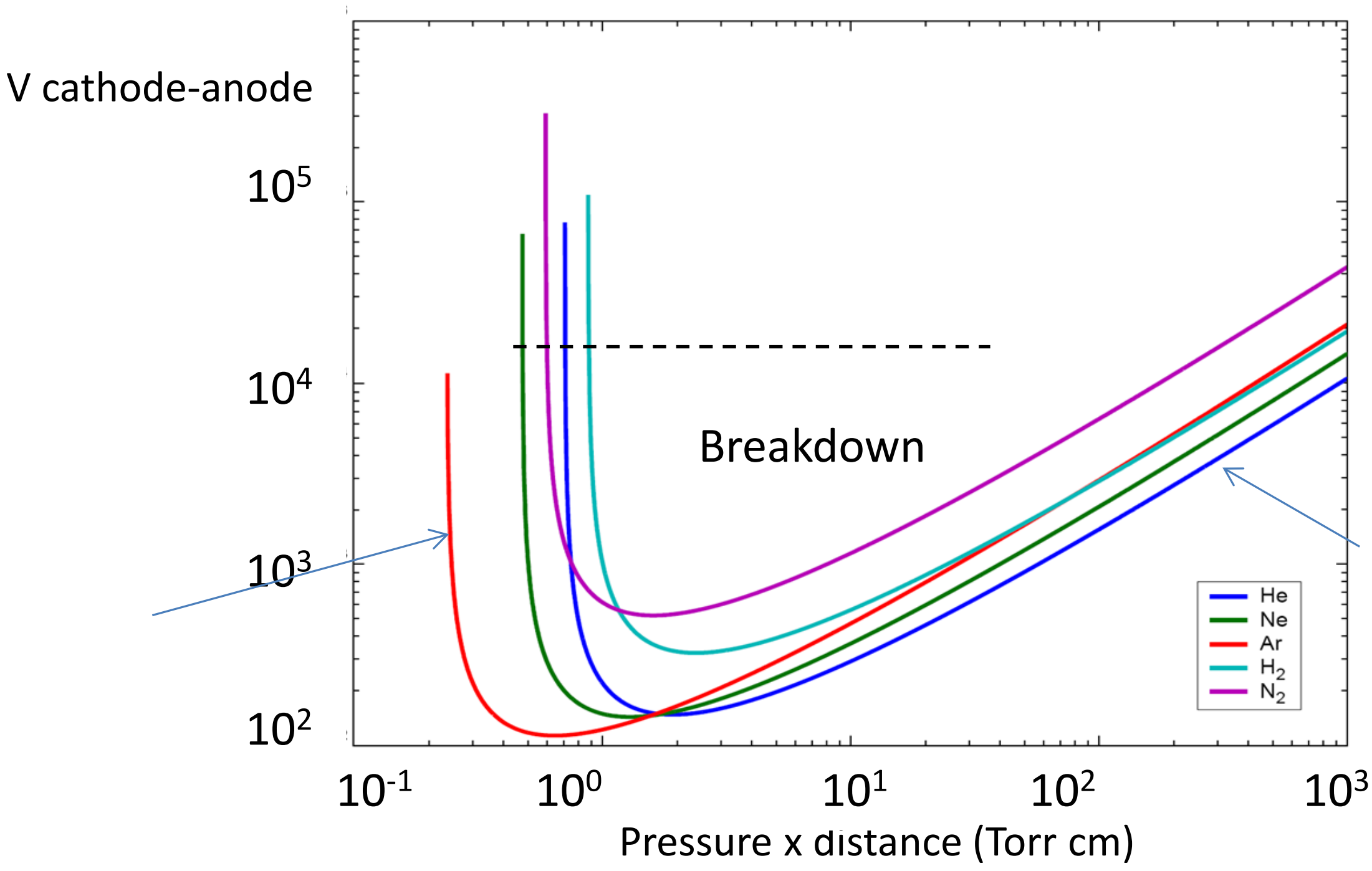

**Fig. 7**: Paschen curves, the lowest voltage for a sustained discharge as a function of the product of gas pressure and anode-to-cathode distance, for various gases.

The point of these electrical discharges is that they contain large quantities of ions that are of interest for ion sources. But to run a gas discharge at low pressures requires a large distance, as the electrons must travel far enough to be able to have an ionizing collision. One way to increase the distance travelled by the electrons is to introduce a magnetic field to the source, and by changing the configuration of the electrodes and magnetic field we find a multitude of different source types, just a few of which are shown with a schematic of the discharge in Fig. 8.

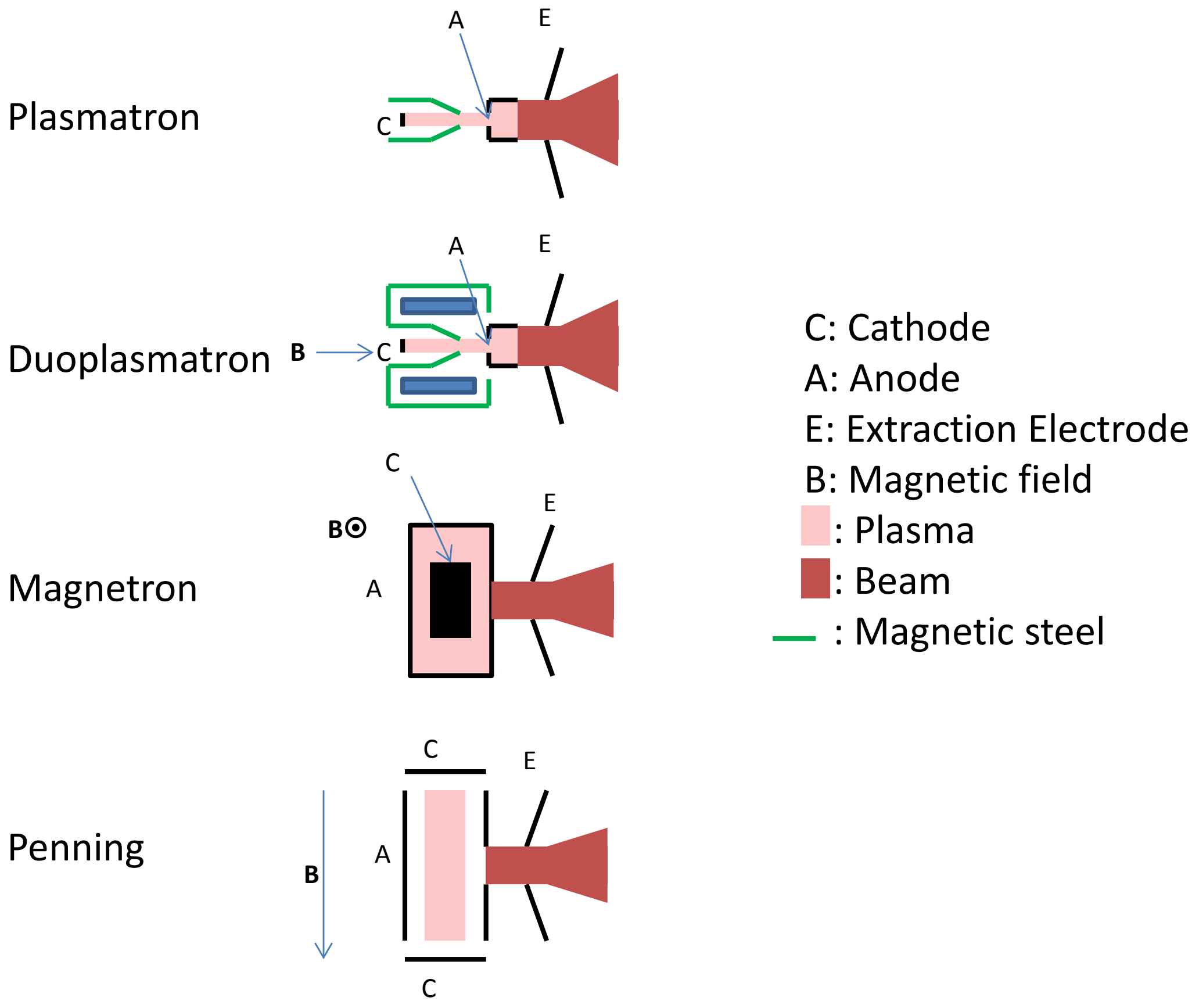

**Fig. 8:** Schematic configuration of a few discharge-type ion sources

Through all these source types there is the possibility to make almost any ion type, with a vast array of intensities and pulse structures, but all sources are limited in their lifetime by the problem that the cathode is constantly bombarded by ions, which sputter away the cathode material. This causes damage to the cathode (changing its surface shape, for example), can cause a change in the cathode composition (either to the crystal structure or to the ratio of elements in a compound cathode), and causes the cathode material to be deposited in places where it is not wanted (for example, across insulators).

The rate at which the cathode is sputtered depends on the cathode material, impacting ion type as well as the energy of impact. Measurements of the sputtering yield (atoms sputtered from a material per incident ion) can be found in the literature (e.g. Ref. [5]). A few plots are reproduced in Fig. 9 as examples, which show how the problem is particularly acute when heavier gases (or vapours) are used.

Part of the solution of the cathode sputtering problem is addressed with radio-frequency (RF), microwave, electron cyclotron resonance (ECR) and laser sources.

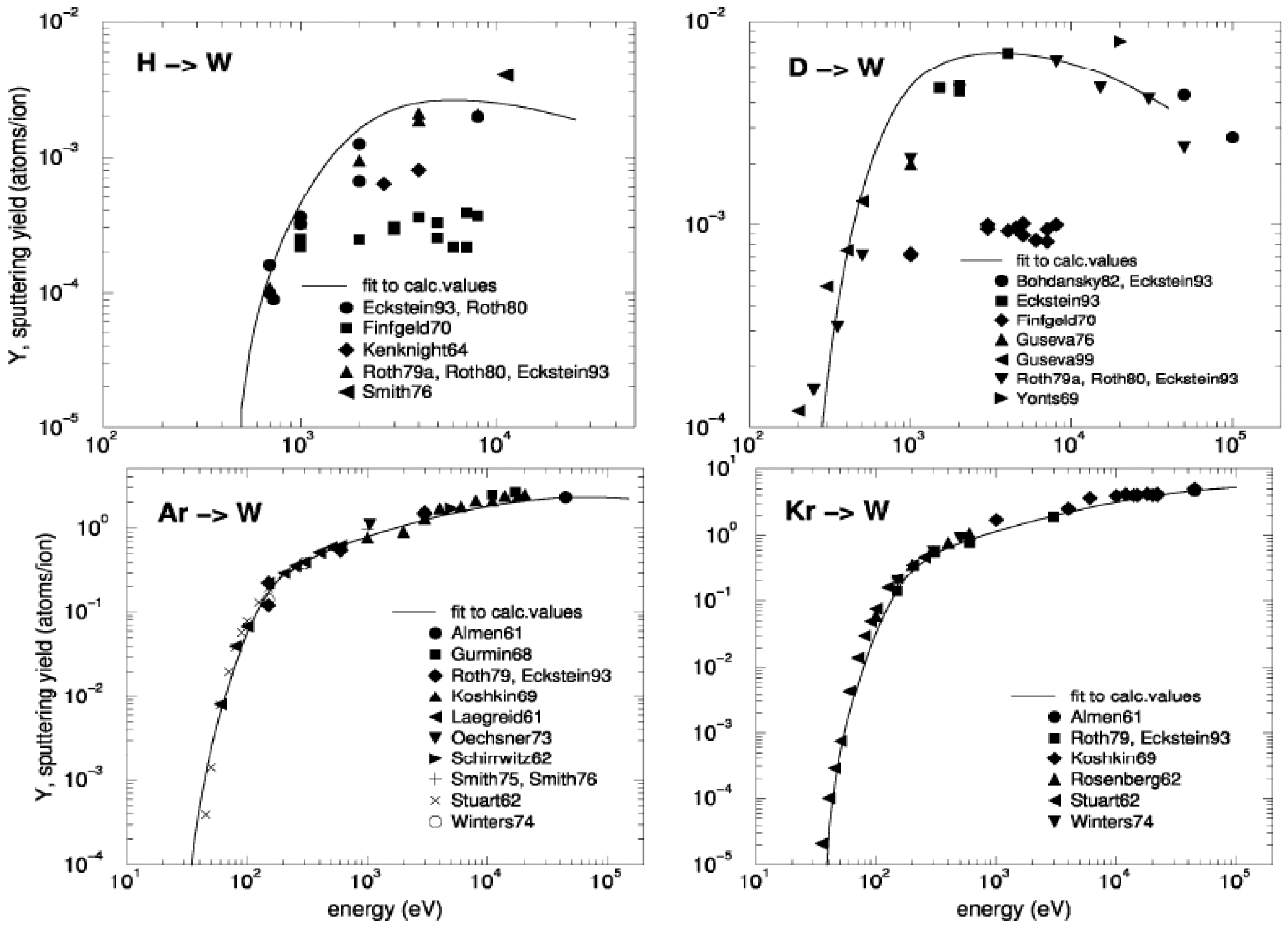

**Fig. 9:** Sputter yield (*Y*) for a few ion types onto tungsten, as a function of energy (from Ref. [5])

### 4.3 RF discharge

Instead of using a cathode–anode voltage to generate the electric field required to accelerate electrons to the energy required to cause ionization, a time-varying electric field can also be used. Two general configurations exist that are able to be applied around a plasma region, and these are to use a capacitive-type system with two electrodes, or a coil to give an inductive system. The two systems are shown schematically in Fig. 10.

Using a capacitively coupled system is conceptually fairly simple, but still has the disadvantage that the electric field lines terminate on the electrodes, which therefore take the place of the anode and cathode, which means that they share the plasma ion load that would normally be taken by the single cathode in a d.c. or pulsed system.

In an inductively coupled system, the electric field is generated by the time variation of the magnetic field, and in the scheme shown in Fig. 10 (right) it has electric field lines that do not terminate on the conductor. Therefore the electric field does not drive ions into the conductor surface and reduces the damage to them.

The electric field is strongest near the coil, and has a strength given by

$$E \approx \frac{\mu_0 \pi N_t I r f}{L}, \tag{12}$$

where $N_t$ is the number of turns, $I$ is the peak current flowing in the coil, $r$ is the coil radius, $f$ is the frequency and $L$ is the coil length. The electric field strength can reach several kilovolts per metre in the right conditions.

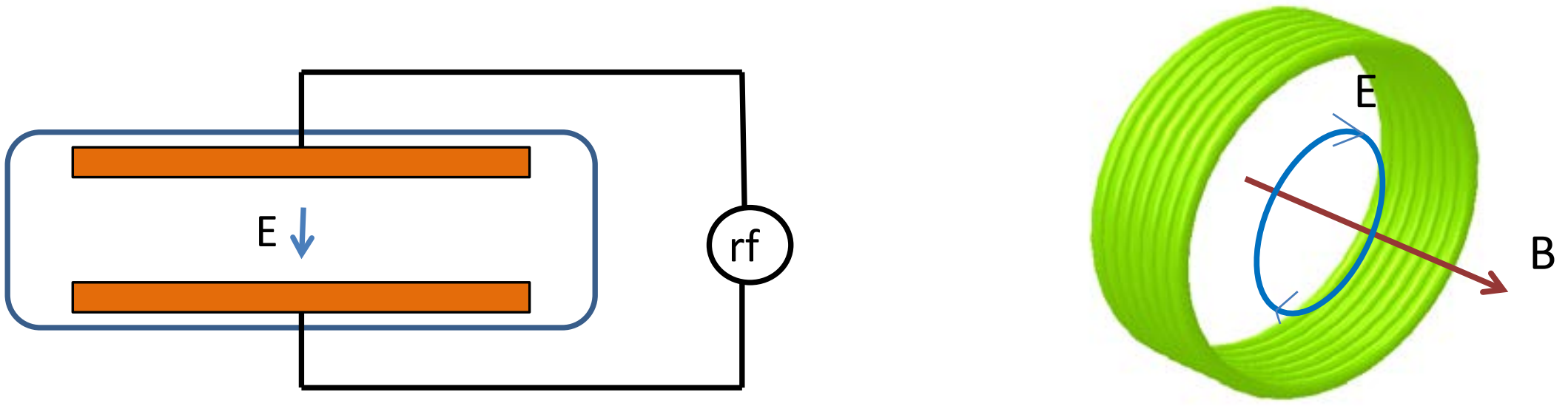

**Fig. 10**: Schematic of two excitation schemes using radio-frequency electromagnetic fields. Left: capacitive coupled system. Right: inductively coupled system.

RF sources typically use relatively low RF frequencies (from hundreds of kilohertz up to a few tens of megahertz) and typically they do not try to couple to any plasma resonance. Therefore, to a first approximation, they behave in a similar way to cathode–anode discharge sources but with a different electron acceleration mechanism.

An example of an RF discharge source using a solenoidal antenna is shown in Fig. 11.

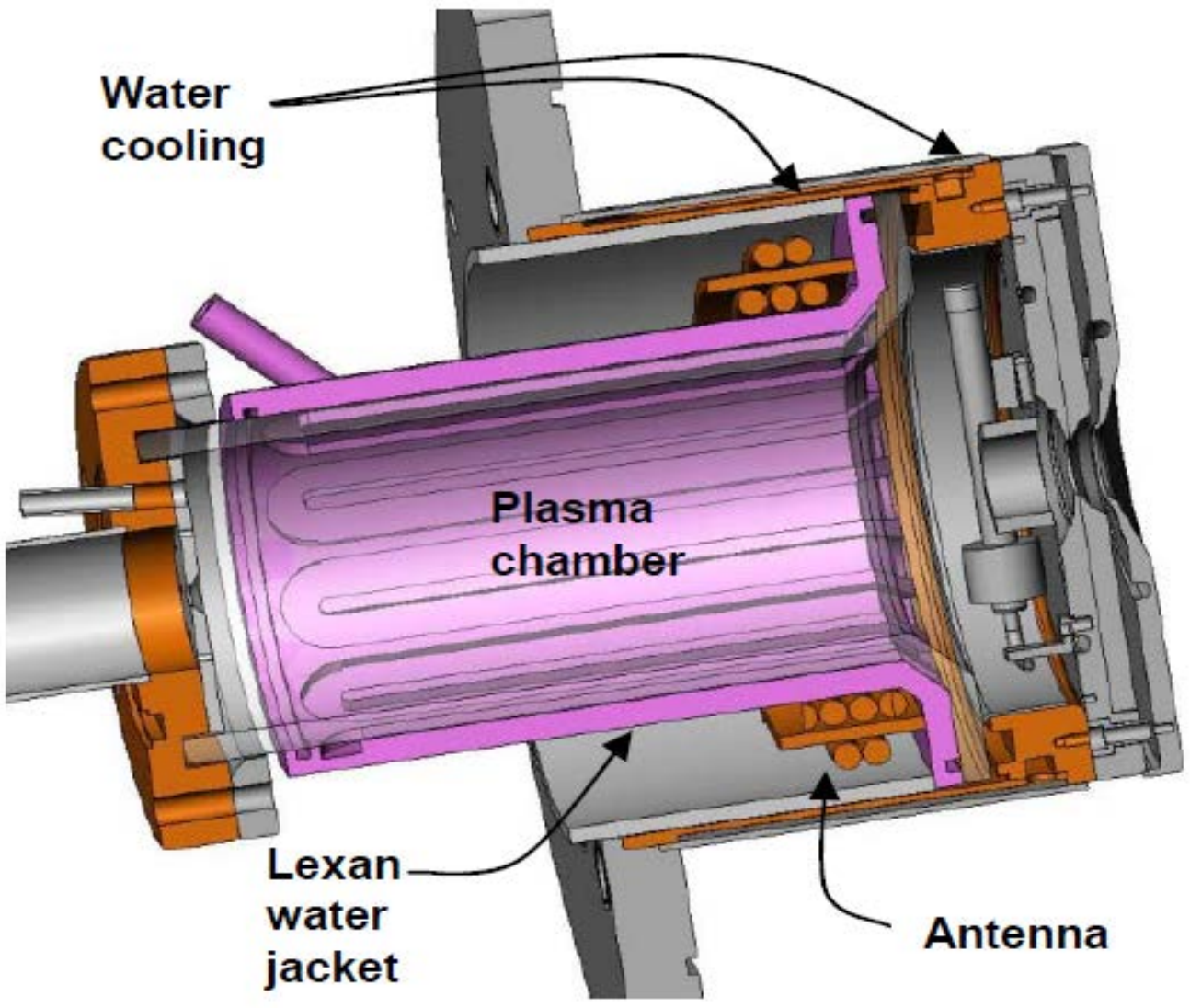

**Fig. 11**: An example of an RF discharge source, from the Spallation Neutron Source, for $H^-$ production

### 4.4 Electron cyclotron resonance ion sources (ECRIS)

We have already learned that charged particles in a magnetic field make circular orbits in a direction perpendicular to the magnetic field, with a frequency ($f_c$) given by

$$f_c = \frac{1}{2\pi}\frac{qeB}{m}, \tag{13}$$

which is independent of the velocity of the particle (as long as relativistic energies are not reached). For electrons, this translates to an engineering formula of $f_c$ = 28 GHz $T^{-1}$. Applying a strong magnetic field to a plasma (for example, 0.5 T with a pair of Helmholtz-type coils) leads to the electrons circulating with a frequency of 14 GHz.

The injection of microwaves of this same frequency into the plasma leads the electric field component of the wave to be in resonance with this electron orbit, and accelerates (or decelerates) the electrons. This coupling of energy heats the electrons very efficiently, to the point where they are able to cause electron impact ionization of the injected gas, or further ionize ions in the plasma.

As they are heated, the process of collisional drift allows the heated electrons to be more efficiently contained in the plasma, which further benefits the creation of energetic electrons.

ECRIS have therefore become a very popular type of ion source, capable of delivering very different types of beams. Lower-frequency versions (typically 2.45 GHz frequency, often using permanent magnets to provide the required 87.5 mT resonance field) are well suited to producing beams of light ions, including high-intensity and high-duty-factor pulsed beams – one such source developed by CEA Saclay [6] is shown in Fig. 12. High-frequency versions are capable of producing beams of high-charge-state ions.

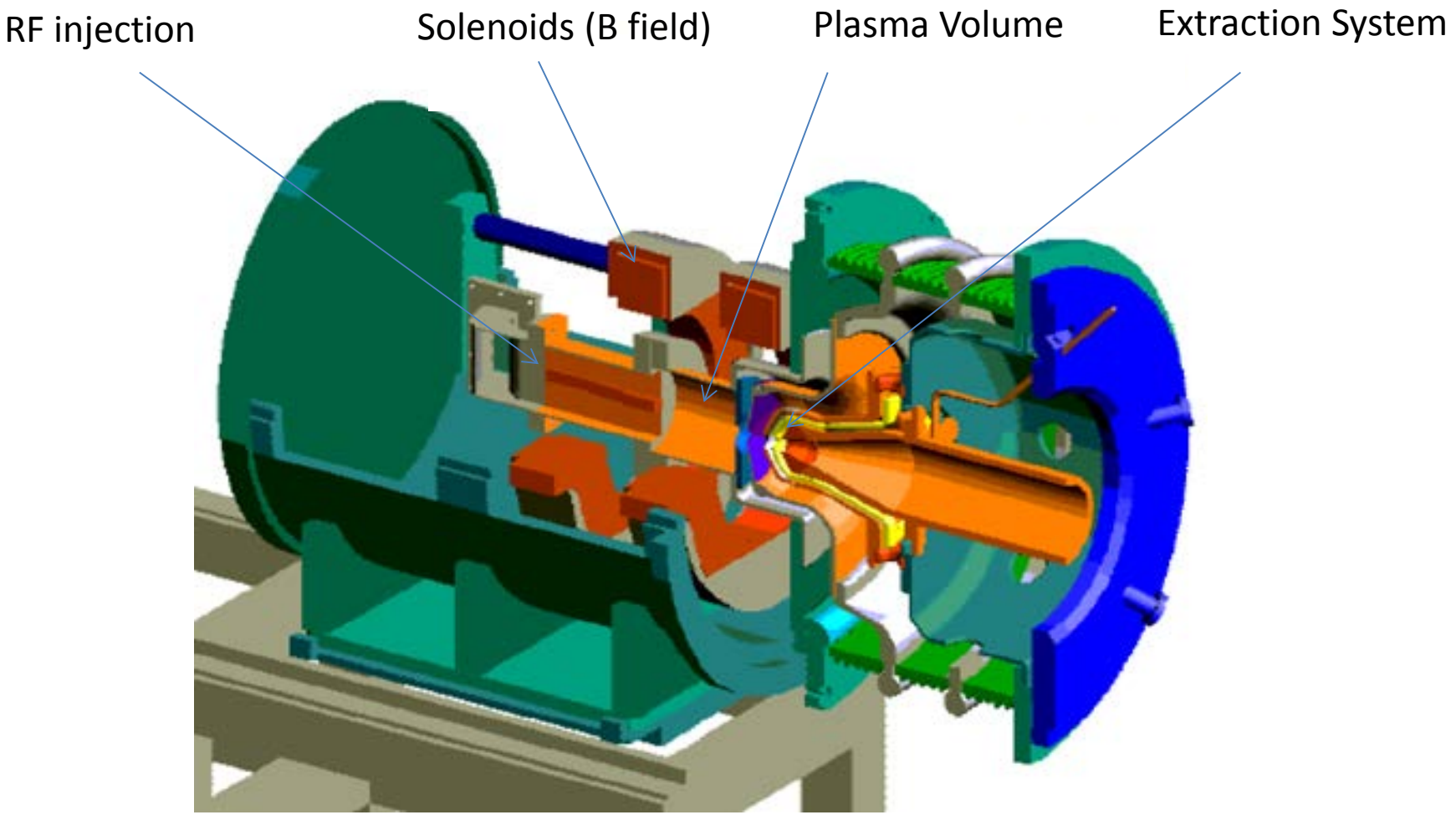

**Fig. 12:** Layout of the SILHI 2.45 GHz ECRIS from CEA

### 4.5 Laser ion sources

Lasers allow us to produce high-brightness, monochromatic beams of photons that can be used in two ways to produce ions, which are explained in the following sections.

#### *4.5.1 Laser ionization ion source*

Photons directed towards atoms, with the photon energy equal to or greater than the ionization potential, can directly ionize the atom. However, calculating the wavelength ($\lambda$) of the light required for this process,

$$\lambda = \frac{hc}{e\Phi_i} = \frac{1.24}{\Phi_i}\ \mu\mathrm{m}, \tag{14}$$

shows that, even for the element with the lowest ionization potential, francium at 3.83 eV, the longest suitable wavelength of 324 nm is already far into the ultraviolet (UV) part of the spectrum, and even this element requires a wavelength only available from a few laser types.

Therefore the ionization process is usually made using multiple photons of different wavelengths, where one beam of laser light raises an electron into an excited state, which then allows the ionization process to take place with a much lower-energy photon, shown schematically in Fig. 13. The use of these element specific energy levels allows a high degree of selectivity of the type of ion produced, as the excitation of atoms between two energy levels requires a very precise photon energy.

In this case, only one element will be put into the excited state, and only these excited atoms will be ionized.

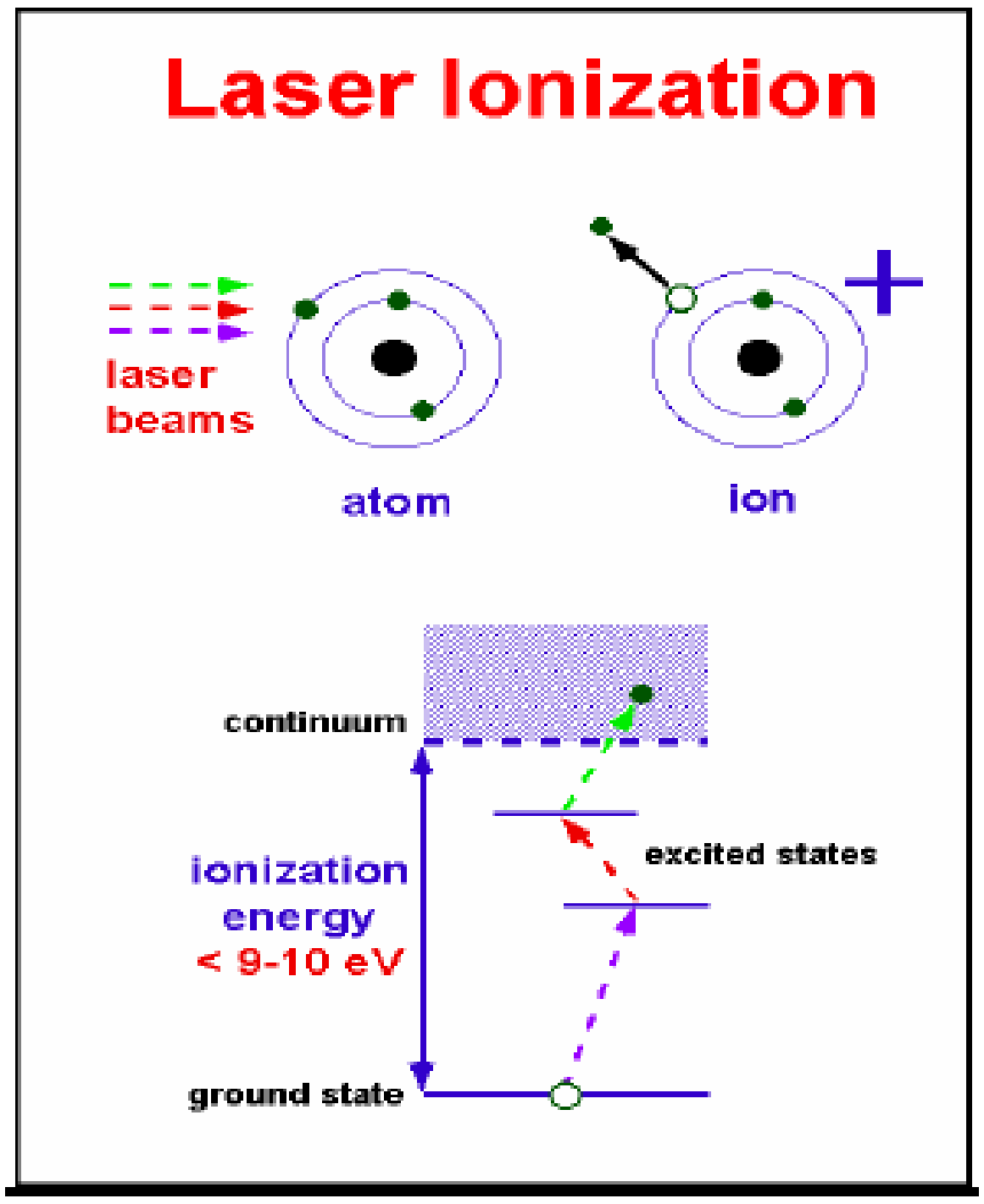

**Fig. 13:** Scheme for ionizing an atom using several laser wavelengths

This type of ion source has therefore lent itself very well to the radioactive ion beam community, where the selectivity based on element type is a complement to the mass spectrometry to allow a very high level of isotope selectivity. This has been taken even further by finding excitation levels whose energy is dependent on the mass of the radioactive nucleus, therefore allowing the isotope selectivity to be made even within the source.

### *4.5.2 Laser plasma ion sources*

With a laser plasma ion source, the intense radiation from a laser is used to produce a dense, and sometimes hot, plasma, in which electron impact ionization is used to produce ions.

As many lasers can be used in a pulsed mode to deliver high instantaneous powers, and the beam of photons can be focused to a very small size, this means that very high power densities can be reached. When directed to a solid surface, these high power densities easily cause the surface to vaporize. In the transition from the density of the solid (approximately $10^{23}$ atoms/cm$^3$) to the low atomic density of vacuum, at some point the density will lead to a plasma electron oscillation frequency that is the equal to the laser frequency.

The relationship between the original laser light wavelength ($\lambda$) and the electron density at which this resonance occurs is given by

$$\lambda = \frac{2\pi c}{e}\sqrt{\frac{m_e \varepsilon_0}{n_e}}, \tag{15}$$

where $m_e$ and $n_e$ are the electron mass and electron number density. This resonant condition is ideal for strongly heating the plasma.

Atoms ablated from the solid surface pass through the plasma region and are ionized. In this case the ion type is changed easily through modification of the target material, allowing a large range of ions to be produced from this source type.

### 4.6 Surface ion sources

Within a solid material, electrons are present with a range of different energy states – Pauli's exclusion principle forces each electron to have a different wavefunction, and therefore a different energy level or spin. This means that electrons populate higher and higher energy levels within the material.

The temperature of the material causes the distribution of the populated energy levels by electrons to be modified, and some electrons to find higher energy levels, which may even allow the electrons to move to a high enough energy to escape the material (and become a thermionic electron emitted from a cathode).

If a layer of atoms coats this material surface, their outer electrons can populate the energy levels of the material; and if the atoms are desorbed from the surface (by heating or bombardment), a fraction will be liberated in a charged state. The Saha–Langmuir equation is used to calculate the ratio of the number of atoms desorbed as ions ($n_\mathrm{i}$) and neutrals ($n_0$) as a function of the base material work function ($\phi_\mathrm{s}$) and the desorbed atom's first ionization potential ($\phi_\mathrm{i}$):

$$\frac{n_\mathrm{i}}{n_0 + n_\mathrm{i}} = \left(1 + \frac{g_0}{g_\mathrm{i}} \mathrm{e}^{(\phi_\mathrm{i} - \phi_\mathrm{s})/kT}\right)^{-1}, \tag{16}$$

where $T$ is the temperature and $g_0/g_\mathrm{i}$ is the ratio of the statistical weights (or degeneracy) of the neutrals to ions. This ratio is plotted in Fig. 14 for two temperatures, and the ratio of desorbed ions is circled on the plot for various elements. It can be seen that ions can still be desorbed even when the ionization potential is higher than the material work function, especially if the material is heated to higher temperatures.

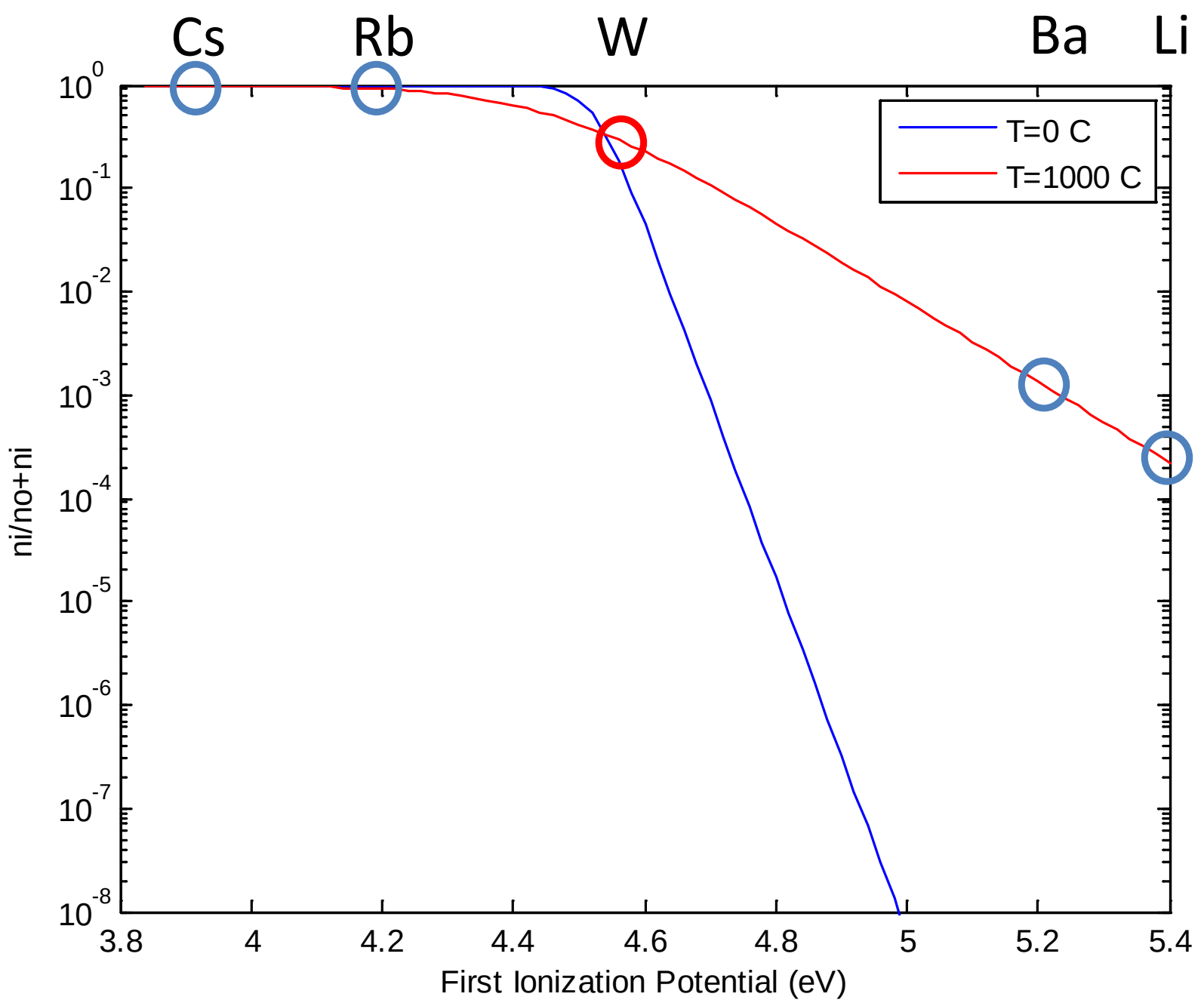

**Fig. 14**: Fraction of particles emitted from a tungsten surface as ions for two different temperatures. The ratios for several different elements are marked.

The design of the source is therefore relatively simple, with a tube of a high-work-function material heated to high temperatures, and injected with the material to be ionized, and is shown schematically in Fig. 15. In some applications, this type of source can be simply a hot filament.

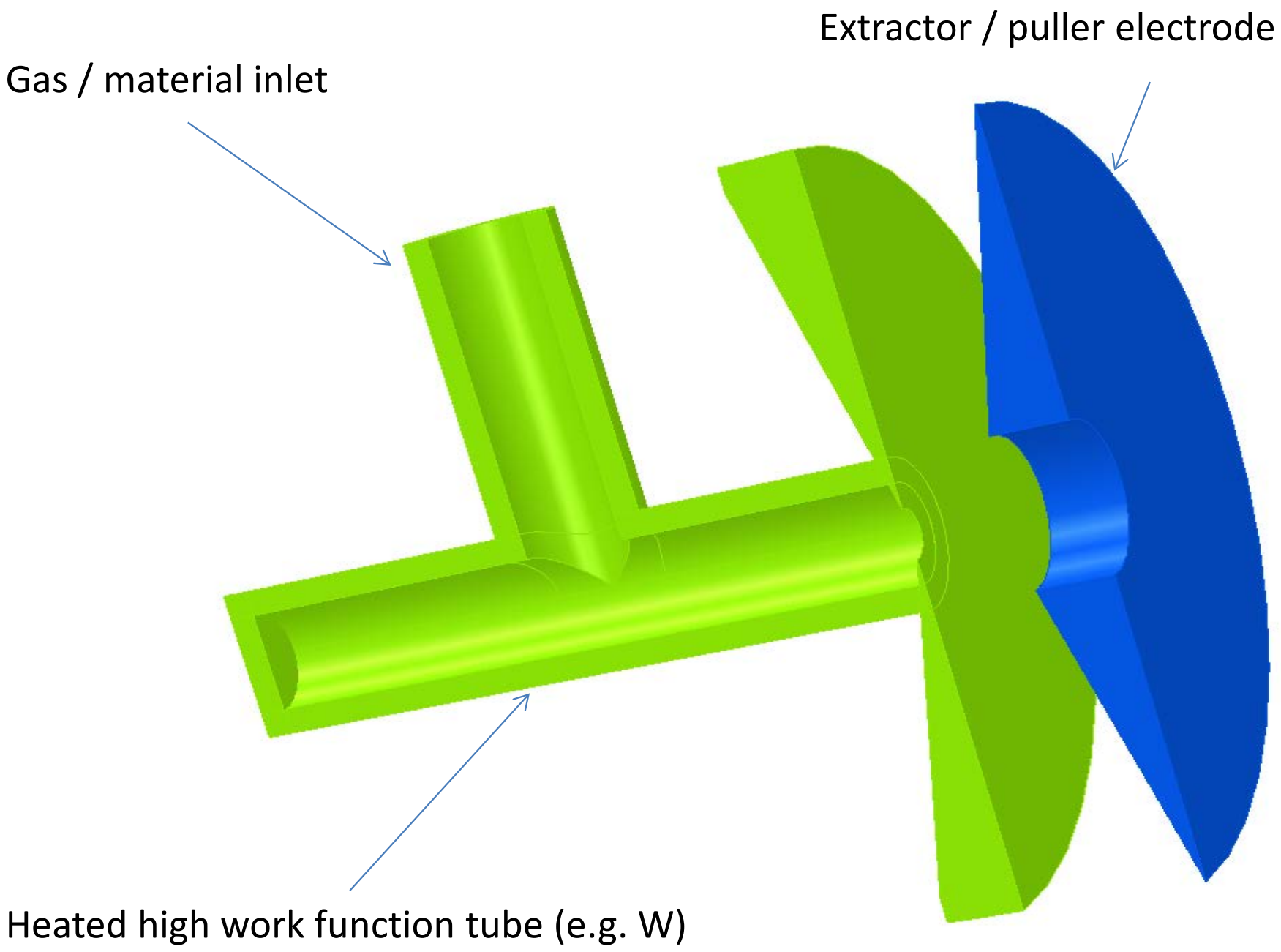

**Fig. 15:** Scheme of a surface ion source

### 4.7 Charge exchange ion sources

Negatively charged ions are possible for a range of elements, especially hydrogen and the halogens (fluorine, chlorine, bromine, etc.), which are said to have a positive electron affinity (i.e., the electron can be bound to the atom, and energy is required to release it). The binding energy is typically small (0.75 eV for $H^-$) and normally the cross-section for destruction is therefore high.

The charge exchange ion source uses the transfer of electrons directly between ions and neutral atoms and molecules. For example, Fig. 16 shows the cross-section for projectile protons ($H^+$) to capture directly two electrons from an $H_2$ molecule as a target, in order to produce $H^-$ ions. This cross-section peaks around 20 keV energy for the projectile $H^+$ ion.

There is no need to have the same type of ion as the projectile and target, and alkali-metal vapours usually provide a much better ratio for electron capture towards the projectile, compared to the cross-section for stripping the electrons off the projectile. Therefore, it can be possible to convert a high fraction (10% or more of some elements, see Fig. 16) into negative ions.

Hence the charge exchange ion source (shown schematically in Fig. 17) firstly requires an ion source for positive ions, which can be of many different types (again depending on the application); it then passes the beam through a charge exchange cell, which should contain the gas or alkali vapour that will serve as the electron donor.

These source types also lend themselves to the production of polarized ions, where the metal vapour ions can be optically polarized, and during the charge transfer this polarization is transferred to the ion and then to the nucleus.

More information about polarized ion sources can be found in Ref. [10].

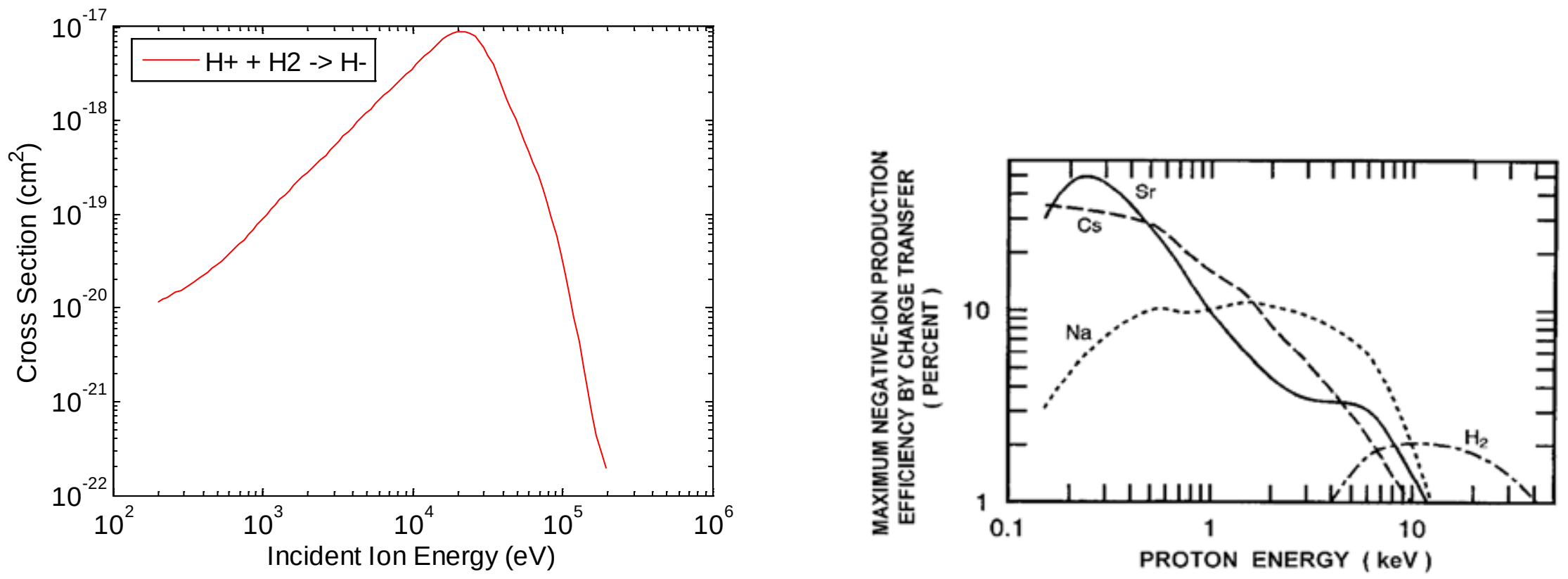

**Fig. 16:** Left: cross-section for double charge transfer of $H^+$ + $H_2$ to $H^-$ ions, as a function of incident $H^+$ energy; data from Ref. [7]. Right: maximum conversion fraction of $H^+$ to $H^-$ ions, as a function of proton energy and target type; reproduced from Ref. [8].

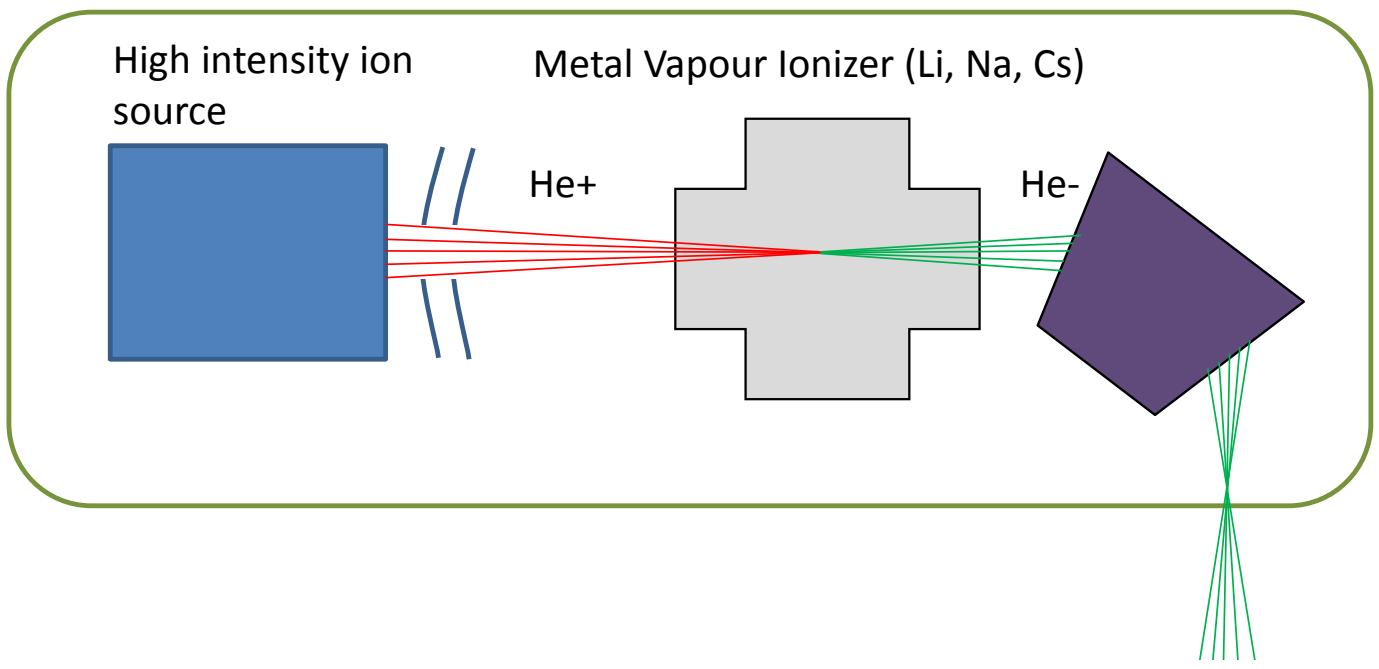

**Fig. 17:** Scheme of a charge exchange-type source; scheme derived from Ref. [9]

## 5 Methods for negative ion production

In section 4.7 it was seen how negative ions can be produced by the transfer of electrons from a gas or vapour to a beam of ions. For hydrogen, there is a range of sources able to deliver negative ions, even with high intensities. They rely on two main physical processes, namely, plasma volume production, and surface production.

The extraction of negative ions is also complicated by the fact that the electric field also extracts electrons, which, although they are accelerated quickly, can be present in sufficient numbers to cause a large increase in the space charge from the source.

### 5.1 Volume production

There are many ways for an additional electron to be attached to a hydrogen atom in a plasma. For example, a free plasma electron can be directly attached; and a charge exchange can occur between two hydrogen atoms, leading to one proton and one $H^-$. But the cross-sections for these processes are small, and, for a given plasma electron temperature, always have a higher cross-section for the destruction of the $H^-$.

There is a particular process where the rate of creation of $H^-$ is higher than the destruction rate, which is by electron attachment of low-energy electrons to an excited $H_2$ molecule, which then disassociates into a neutral H and an $H^-$. In order to create the excited molecules, a higher electron temperature is needed to create the excited $H_2$, but this zone can be separated from the $H^-$ production area by a magnetic field, which as a result of collisional drift (see section 3.3) will allow lower-energy electrons to pass more quickly.

The plasma is produced and heated by several possible methods, including low-frequency RF as well as cathode and gas discharges (for example, in Penning sources).

### 5.2 Surface production

The production of positive ions from surfaces was discussed in section 4.7, where it was explained that an electron can be captured by a surface as an atom is desorbed from the surface. If the work function of the surface is very small, it is also possible that a desorbed atom has an additional electron attached.

For the best effect, the work function of the surface should be close to the binding energy of the electron to the atom, which for hydrogen would mean using a work function close to 0.75 eV. A coating of an alkali metal – the best being caesium – on a metal surface is the way to produce some of the lowest work functions, particularly if the layer is very thin (for the lowest possible work function, the layer of Cs should be less than one atom thick). A hydrogen plasma is then used to coat the caesium layer with hydrogen, and then bombard it with ions, which desorb these hydrogen atoms from the surface, with a high probability of them being negatively charged.

For consistent operating conditions, the source requires that the surface work function be kept reasonably constant, which is a challenge in the environment of an ion source. In addition, the heavy caesium atoms are themselves easily ionized if they are in the plasma, and they can cause significant sputtering of any cathode electrodes in the source.

Using caesium in the ion source usually leads to increased intensities, as well as decreasing the density of electrons at the extraction region, reducing the intensity of this co-extracted beam.

### 5.3 Highly charged ions

Highly charged ions are of interest for accelerators because they allow more energy gain for the same electric field. This is most interesting when the beam is of low intensity, where only a small fraction of the power put into an accelerator is transferred to the beam itself.

The production of highly charged ions cannot be normally made in a single step ($A^0 \rightarrow A^{q+}$), either by electron impact ionization or photoionization, as the electron or photon will need to have enough energy to liberate all the atomic binding energies. So logically the method is to continue to bombard ions with electrons, removing one electron at a time, called stepwise ionization. In this case the highest electron energy needs only be sufficient to cause the ionization of the highest charge state, and the ions need to be kept in the plasma sufficiently long for them to have enough collisions to allow them to reach the desired charge state.

The ionization potentials for all ions and charge states have been calculated by Carlson, Nestor, Wasserman and McDowell [11], and the cross-sections for ionization by electron impact from a charge state $q$ to $q + 1$ ($\sigma_{q\rightarrow q+1}$) can be estimated for all charge states and ions from Lotz's formula [12].

As the electron energy requirements are therefore known to reach the highest charge state, and the cross-sections can then be estimated, we can use the following formula to link to the electron density ($n_e$) and the time required to reach the wanted charge state ($\tau$):

$$n_e\tau = \frac{1}{v_e}\sum\frac{1}{\sigma_{q\rightarrow q+1}}, \tag{17}$$

where $v_e$ is the average electron velocity. We see that it is the product of $n_e \tau$ that is important, i.e., we need high electron densities or a long time to reach the high charge states.

The three sources that generally fit these requirements for producing highly charged ions (i.e., ions where the last ionization potential is a few hundred volts) are the following:

- *Electron cyclotron resonance ion sources.* Electrons can be heated to high energies, and ions are trapped in the potential well of the electrons for many milliseconds.
- *Electron beam ion sources.* Electrons are produced by a cathode, the gun accelerates them to the energy required for the highest ionization, and this beam is focused. Ions are trapped by the potential of the electron beam, as well as electric fields applied by electrodes.
- *Laser plasma ion sources.* A very hot dense plasma is created by laser heating of a surface and then the plasma. Ions travel through the plasma quickly, but, owing to the high electron density in the plasma, they can undergo sufficient collisions to become highly ionized within a few nanoseconds.

It should not be forgotten that the counterpart of ionization is recombination, for which the cross-sections increase as the ion charge state increases. Of particular importance are recombination through charge exchange with other ions and atoms (avoided by keeping the background pressure low in the source) and three-body recombination, where one electron is recombined while another takes away the excess energy. This three-body recombination generally works against very dense plasma sources.

## 6 Conclusions

Within this chapter we discovered some of the fundamental processes that allow the production of ionized atoms, and saw how they are applied to ion sources. Depending on the main mechanism used for producing the ion, we find that there are different classifications of ion sources.

Ultimately the application should define the type of ion source used, where the source will be chosen that has the best compromise for plasma properties for the desired ion beam properties, as well as the intensity and pulse characteristics, simplicity of implementation and cost.